\documentclass{article}

\usepackage{arxiv}
\usepackage{lipsum} % Required to insert dummy text
\usepackage[version=4]{mhchem}
\usepackage{siunitx}
\DeclareSIUnit\Molar{M}
\usepackage{graphicx}
\usepackage{hyperref}
\usepackage{amsfonts}
\usepackage{amsbsy}
\usepackage{amssymb}
\usepackage{xcolor}
\usepackage[normalem]{ulem}
\usepackage{float}
\usepackage{multirow}
\usepackage{ragged2e}
\usepackage[utf8]{inputenc} % allow utf-8 input
\usepackage[T1]{fontenc}    % use 8-bit T1 fonts
\usepackage{hyperref}       % hyperlinks
\usepackage{url}            % simple URL typesetting
\usepackage{booktabs}       % professional-quality tables
\usepackage{amsfonts}       % blackboard math symbols
\usepackage{nicefrac}       % compact symbols for 1/2, etc.
\usepackage{microtype}      % microtypography
\makeatletter
\def\blfootnote{\xdef\@thefnmark{}\@footnotetext}
\makeatother
\usepackage{graphicx}
\graphicspath{ {./images/} }

\title{Neural complexity --  Statistical-mechanical approach of human electroencephalograms}

\author{
   \noindent\large{Dimitri Marques Abramov\textit{$^{a}$}, Constantino Tsallis\textit{$^{b}$}, and Henrique Santos Lima\textit{$^c$}}
}

\begin{document}
\maketitle

\begin{abstract}
The brain is a complex system whose understanding enables potentially deeper approaches to mental phenomena. Dynamics of wide classes of complex systems have been satisfactorily described within $q$-statistics, a current generalization of Boltzmann-Gibbs (BG) statistics. Here, we study human electroencephalograms of typical human adults (EEG), very specifically their inter-occurrence times across an arbitrarily chosen threshold of the signal (observed, for instance, at the midparietal location in scalp). The distributions of these inter-occurrence times differ from those usually emerging within BG statistical mechanics. They are instead well approached within the $q$-statistical theory, based on non-additive entropies characterized by the index $q$. The present method points towards  a suitable tool for quantitatively accessing brain complexity, thus potentially opening useful studies of the  properties of both typical and altered brain physiology. 
\end{abstract}
% \begin{document}
% \flushbottom

% \maketitle
% \dates{This manuscript was compiled on \today}
\blfootnote{\textit{$^{a}$~Instituto Nacional da Saúde da Crianca, da Mulher e do Adolescente Fernandes Figueira. Fundação Oswaldo Cruz, Avenida Rui Barbosa 716, Flamengo, Rio de Janeiro 22250-020, Brazil.\\ 
E-mail: dimitri.abramov@iff.fiocruz.br }}

\blfootnote{\textit{$^{b}$~Centro Brasileiro de Pesquisas Fisicas and National Institute of Science and Technology of Complex Systems, Rua Xavier Sigaud 150, Rio de Janeiro-RJ 22290-180, Brazil \\
%Fax: XX XXXX XXXX; Tel: XX XXXX XXXX; 
Santa Fe Institute, 1399 Hyde Park Road, Santa Fe, 
 New Mexico 87501, USA \\
Complexity Science Hub Vienna, Josefst\"adter Strasse 
 39, 1080 Vienna, Austria \\
E-mail: tsallis@cbpf.br}}

\blfootnote{\textit{$^{c}$~Centro Brasileiro de Pesquisas Fisicas, Rua Xavier Sigaud 150, Rio de Janeiro-RJ 22290-180, Brazil.\\ 
E-mail: hslima94@cbpf.br }}

% \thispagestyle{firststyle}
% \ifthenelse{\boolean{shortarticle}}{\ifthenelse{\boolean{singlecolumn}}{\abscontentformatted}{\abscontent}}{}
\section{Introduction
}
The brain is widely recognized as a complex system since it is composed by billions of cells (neurons) which express individual behaviors and, at same time, they build a fully interconnected network with emergent, self-organized collective behaviors \cite{BassetGazzaniga2011}. Thus, traditional reductionist scientific methodology from mechanistic rationality appears to fail for deeply understanding the brain and its associated mind 
inside a multidimensional environment 
\cite{Braslow2021}. On one hand, a humanity's great unresolved problem is to establish a suitable mental medicine, from epistemology \cite{Canguilhem1991} to the biomedical perspective. The problem begins in differentiating normality from typicality, illness from neurodiversity. And, upon this basis, to establish a taxonomy about mental typology for a more realistic nosography. On the other hand, several studies have explored brain complexity through entropic measures within the electroencephalogram (EEG), and found relationships between brain complexity and different mind conditions \cite{Lauetal2022}. However, this issue yet is incipient. 

The pioneering works of Boltzmann \cite{Boltzmann1872} and Gibbs \cite{Gibbs1901} (BG) established a magnificent theory which is structurally associated with the BG entropic functional 
\begin{equation}
%\label{BGentropy}
S_{BG}=-k\sum_{i=1}^W p_i \ln p_i \;\;(\sum_{i=1}^W p_i=1)\,,
\label{BGentropy}
\end{equation} 
and consistent expressions for continuous or quantum variables; $k$ is a conventional positive constant adopted once forever (in physics, $k$ is chosen to be the Boltzmann constant $k_B$; in information theory and computational sciences, $k=1$ is frequently adopted). 

In the simple case of equal probabilities, this functional becomes $S_{BG}=k \ln W$. 
Eq. (\ref{BGentropy}) is generically {\it additive} \cite{Penrose1970}. Indeed, if $A$ and $B$ are two probabilistically independent systems (i.e., $p_{ij}^{A+B}= p_i^A p_j^B$), we straightforwardly verify that $S_{BG}(A+B)=S_{BG}(A)+S_{BG}(B)$. The celebrated entropic functional (\ref{BGentropy}) is consistent with thermodynamics for all systems whose $N$ elements are either independent or weakly interacting in the sense that only basically local (in space/time) correlations are involved. For example, if we have equal probabilities and the system is such that the number of accessible microscopic configurations is given by $W(N) \propto \mu^N\; (\mu>1; \,N\to\infty)$, then $S_{BG}(N)$ is {\it extensive} as required by thermodynamics. 
Indeed $S_{BG}(N)=k\ln W(N) \sim k(\ln \mu)N$. 

However, complex systems are typically composed of many elements which essentially are non-locally correlated, building an intricate network of interdependencies from where collective states can emerge \cite{GellMannTsallis2004}. BG statistical mechanics appears to be generically inadequate for such systems because this theory assumes (quasi) independent components with short-range (stochastic or deterministic) interactions.   

Indeed, if the correlations are nonlocal in space/time, $S_{BG}$ may become thermodynamically inadmissible. Such is the case of equal probabilities with say $W(N) \propto N^\nu \;(\nu>0; \,N\to\infty)$: it immediately follows $S_{BG}(N) \propto \ln N$, which violates thermodynamical extensivity \cite{GellMannTsallis2004}. 
To satisfactorily approach cases such as this one, it was proposed in 1988 \cite{Tsallis1988} 
%\cite{Tsallis1988,TsallisMendesPlastino1998,GellMannTsallis2004,Tsallis2009}
to build a more general statistical mechanics based on the {\it nonadditive} entropic functional
\begin{equation}
S_q\equiv k\frac{1-\sum_{i=1}^W p_i^q}{q-1}=k\sum_{i=1}^W p_i \ln_q \frac{1}{p_i} = -k\sum_{i=1}^W p_i^q \ln_q p_i = -k\sum_{i=1}^W p_i \ln_{2-q} p_i \;\;(q \in \mathbb{R}; S_1=S_{BG})\,,
\end{equation}
with the $q$-logarithmic function $\ln_q z \equiv \frac{z^{1-q}-1}{1-q} \; (\ln_1 z=\ln z)$, its inverse being the $q$-exponential  $e_q^z \equiv [1+(1-q)z]_{+}^{1/(1-q)}$; $(e_1^z=e^z$; $[z]_{+}=z$ if $z>0$ and vanishes otherwise); for $q<0$, it is necessary to exclude from the sum the terms with vanishing $p_i$. We easily verify that equal probabilities yield $S_q=k\ln_q W$. Also, we generically have the following functional nonadditivity
\begin{equation}
\frac{S_q(A+B)}{k}=\frac{S_q(A)}{k}+\frac{S_q(B)}{k}+(1-q)\frac{S_q(A)}{k}\frac{S_q(B)}{k} \,.
\end{equation}
%hence
%\begin{equation}
%S_q(A+B)=S_q(A)+S_q(B)+\frac{1-q}%{k}S_q(A)S_q(B) \,.
%\end{equation}
Consequently, in the $(1-q)/k \to 0$ limit, we recover the $S_{BG}$ additivity. For the anomalous class of systems mentioned above, namely if $W(N) \propto N^\nu$, we obtain, $\forall \nu$,  the {\it extensive} entropy $S_{1-1/\nu}(N)=k\ln_{1-1/\nu}W(N) \propto N$, as required by the Legendre structure of thermodynamics \cite{Tsallis2022,
%,TsallisCirto2013
Tsallis2009}. Finally, the optimization of $S_q$ under simple constraints yields $q$-exponential distributions for the (quasi)stationary states, instead of the usual BG exponentials.
\begin{figure}[htb]
\centering
\includegraphics[scale=0.50,angle=0]{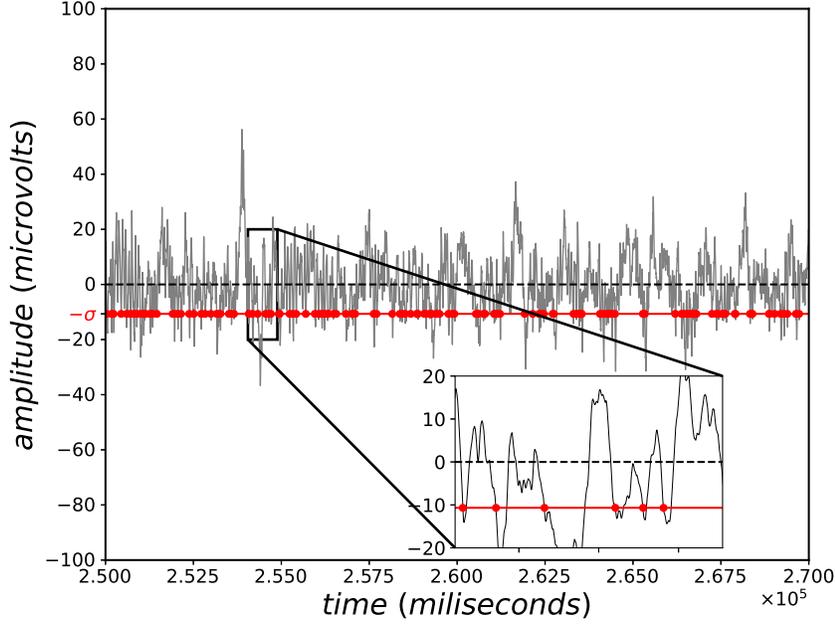}
\vspace{-0 cm} %{2.5}
\caption{\small Segment of ongoing EEG from one subject (B006), recorded on the mid-parietal ($P_z$) location of the head. Red dots:  time values when ddp (signal amplitude) crosses downwards the bottom threshold (1.0 standard deviation; red line). EEG sampling rate was 1000 Hz. 
%{\it Inset:} detail.
} 
\label{fig1} 
\end{figure}

\begin{figure}[htb]
\centering
\includegraphics[scale=0.50,angle=0]{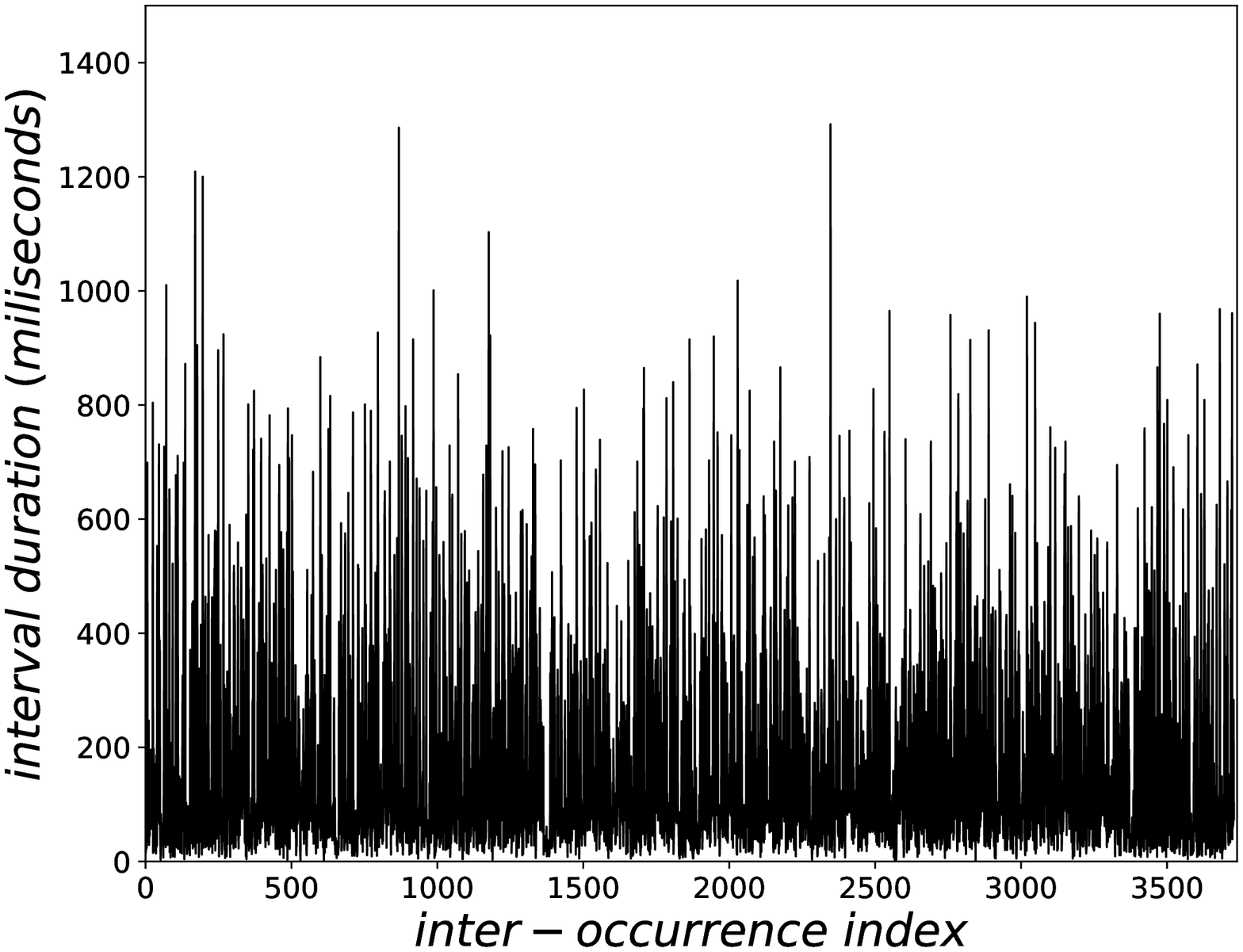}
\vspace{-0 cm} %{2.5}
\caption{\small Sequence of inter-event time intervals from EEG signal, as detected in FIG.~\ref{fig1}.  
%{\it Inset:} detail.
} 
\label{fig2} 
\end{figure}

\section*{Methodology and results}
We analized the EEG signal of ten typical adult humans from a match-to-sample task experiment with neutral affective interference for access working memory and attention, such in Yang and Zhen's study \cite{YangZhen2021}. This work was approved by our ethical board for human research, under CAAE 50137721.4.0000.5269. Each EEG signal has 5-10 minutes length recorded with open eyes at 1000Hz sampling rate, through 20 channels disposed at 10-20 montage with eyes open. The high, low and band-pass filters were respectively 0.5, 150Hz and 60Hz. We did not apply any other filter to minimize signal manipulation. 

\begin{figure}[htb]
\centering
\includegraphics[scale=0.5,angle=0]{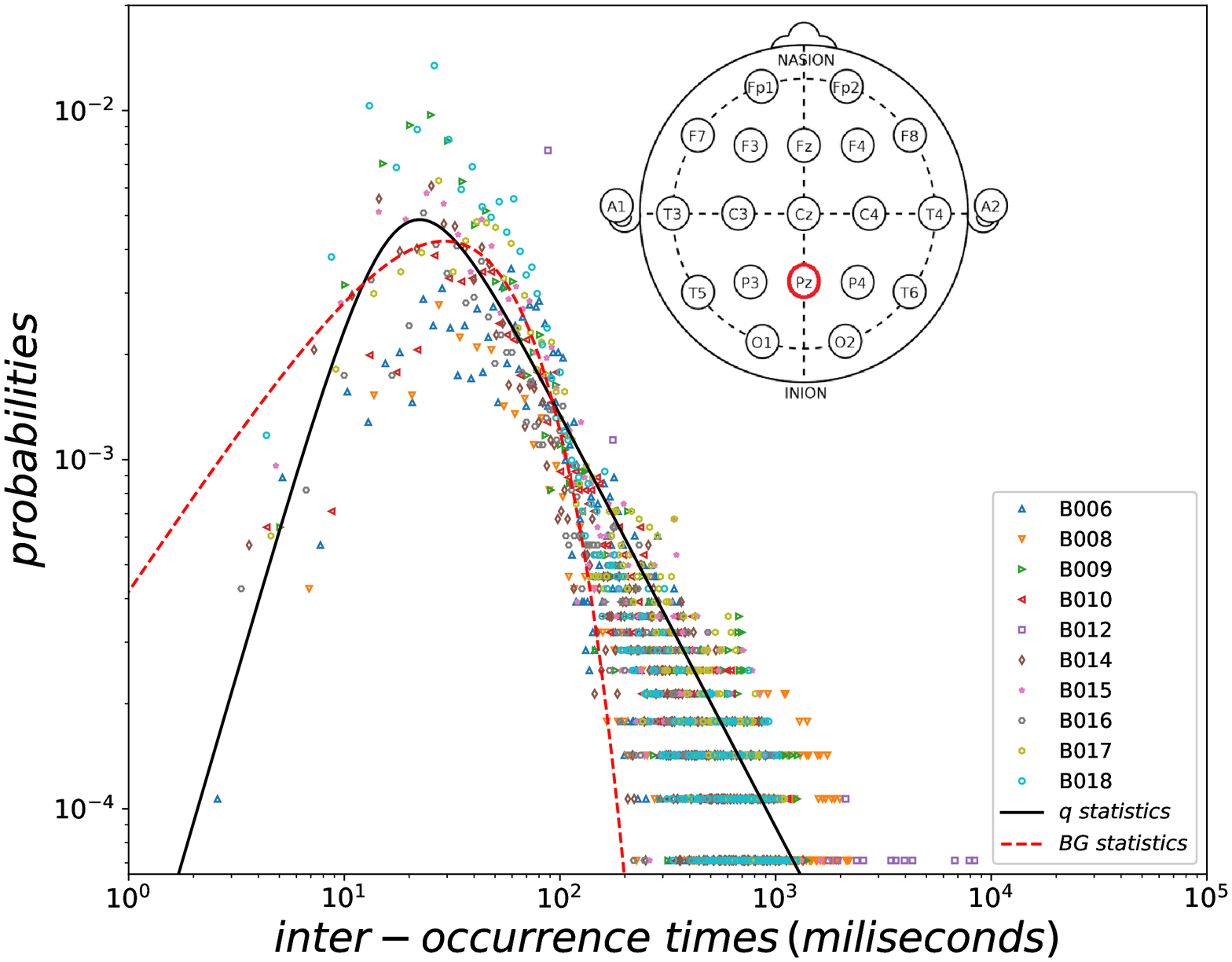}
\caption{\small Probability distributions of EEG inter-occurrence times (500 equal logarithmic bins) and fittings with statistical models. Superimposed signal recorded on the $P_z$ location of ten subjects performing a work memory task.
%(see legend). 
Amplitude threshold = 1.0 standard deviation. Fitting within Boltzmann-Gibbs statistical mechanics for non-complex systems (i.e., $q=1$, dashed red curve).
%with $y_{BG}=a_{BG}\,x^{c_{BG}}\,e^{-\beta_{BG} \,x^{\eta_{BG}}}$ [$(a_{BG},b_{BG},c_{BG},\eta_{BG},q_{BG})=(4.3\times 10^{-4},0.023,0.94,0.93,1)$; dashed red curve]. 
Fitting within nonextensive statistical mechanics for complex systems (i.e., $q\ne 1$, black continuous curve). See Methodology for details.
%with $y_q=a_q\,x^{c_q}\,e_q^{-\beta_q\,x^{\eta_q}}=
%a_q\,x^{c_q}/[1+(q-1)\beta_q \,x^{\eta_q}]^{\frac{1}{q-1}}$ [$(a_{q},b_{q},c_{q},\eta_{q},q)=(2.1\times 10^{-5},2\times 10^{-5},2.12,2.96,1.897)$; black continuous curve]. The  fittings for $q$-statistics were done through  {\it dog leg trust region} \cite{Powell1970} and {\it trust region reflective} \cite{Coleman1996} algorithms, and the results coincide. The fittings for BG-statistics were done through  {\it dog leg trust region}, {\it trust region reflective} and {\it crow search} \cite{Askar2016}) algorithms, and the results once again coincide.
}
\label{fig3} 
\end{figure}

We accessed signal recorded at the midparietal ($P_z$) site (see FIG.~\ref{fig3}), where classical cognitive event-related potentials, as P300 \cite{Polich2006}, manifest during attention tasks. A threshold was set at -1.0 standard deviation from $P_z$ signal average (FIG.~\ref{fig1}, from subject B006). Taking negative voltages we are minimizing the effect of blink artifacts, which are positive waves, amplier in frontal places. 

Each event is the numerical position {\it i} of signal vector (1 second = 1000 positions) where the amplitude crossed the threshold downwards. The inter-event distances $i_n - i_{n-1}$ (where n = 1,...,N) were calculated (FIG.~\ref{fig2}, from B006). The logarithm distribution of inter-event distances (with 500 distance classes) of all ten EEG signals at $P_z$ were superimposed, and the fitting was performed to the following $q$-statistical function (FIG.~\ref{fig3}):
\begin{equation}
y_q=a_q\,x^{\,c_q}/[1+(q-1)\beta_q \,x^{\,\eta_q}]^{\frac{1}{q-1}}\,,
\end{equation}
where $(a_{q},\beta_{q},c_{q},\eta_{q},q)=(2.1 \times 10^{-5},2.0 \times 10^{-5},2.12,2.96,1.89)$ for the best fitting. And, for comparison, we also included the classical statistical BG function (where $q=1$), as follows:
\begin{equation}
y_{BG}=a_{BG}\,x^{\,c_{BG}}\,e^{-\beta_{BG} \,x^{\,\eta_{BG}}}\,.
\end{equation}
where $(a_{BG},\beta_{BG},c_{BG},\eta_{BG},q_{BG})=(4.3 \times 10^{-4},0.023,0.94,0.93,1)$ for the best fitting.

The fitting was performed using three different methods:  dog leg trust region~\cite{Powell1970}, trust region reflective~\cite{Coleman1996}  and crow search~\cite{Askar2016} algorithms, all available in Scipy library. 

The constant $a$ is determined by imposing normalization, i.e., $\int_0^\infty dx\,y(x)=1$. Consequently,
\begin{equation}  a_q^{-1}=\int_0^{\infty} dx\,\frac{x^{c_q}}{[1+(q-1)\beta_q x^{\eta_q}]^{\frac{1}{q-1}}}=(b_q(q-1))^{-\frac{c_q+1}{\eta_q}}\frac{\Gamma(\frac{1+c_q}{\eta_q})\Gamma(\frac{1}{q-1}-\frac{1+c_q}{\eta_q})}{\eta_q\Gamma(\frac{1}{q-1})}
\end{equation}
for $q>1$ and $\frac{1}{q-1}-\frac{1+c_q}{\eta_q}>0$. In the $q\to1$ limit, we obtain
\begin{equation}
   a_{BG}^{-1}= \frac{\beta_{BG}^{-\frac{c_{BG}+1}{\eta_{BG}}} \Gamma \left(\frac{c_{BG}+1}{\eta_{BG}}\right)}{\eta_{BG}}.
\end{equation}
It is observed that EEGs at $P_z$ position from all subjects express very similar distributions of distances. The EEG regularity was modelled by the $q$-statistics function instead BG one (FIG.~\ref{fig3}).

%Through all these examples,  

\section*{Discussion}
This preliminary study exhibits
%seems to be a proof of concept 
that $q$-statistics can somehow reveal the brain complexity, at least for typical people on scalp regions where cognitive-related potential P300 emerges (a site almost free of blink artifacts). Consistently, we have verified here that the brain phenomenology is not properly described within BG statistics (i.e., $q$=1).This is by no means surprising since BG statistics generically disregards inter-component long-range correlations and their collective behavior, which is well known in neural systems \cite{BassetGazzaniga2011}. In contrast, plethoric evidence exists that $q$-statistics satisfactorily models vast classes of complex systems \cite{Tsallis2009} even involving $c_q \ne 0$, from basic chemical reactions through quantum tunneling \cite{Wildetal2023} to financial market behavior \cite{TsallisAnteneodoBorlandOsorio2003}, COVID-19 spreading \cite{TsallisUgur2020}, commercial air traffic networks \cite{Mitsokapasetal2021} (see \cite{LudescherTsallisBunde2011,LutzRenzoni2013,BogachevKayumovBunde2014,CombeRichefeuStasiakAtman2015,GrecoTsallisRapisardaPluchinoFicheraContrafatto2020} for illustrations with $c_q = 0$). We are led to believe that we are dealing with universality classes of complexity, thus revealing, in what concerns information processing and energy dynamics, far more integrative networks than one might a priori expect from neural structures \cite{Sauseng2010}.

By generalizing the BG theory, $q$-statistics shows that it could be a most suitable and promising path to explore brain complexity. Our expectancy is that the $q$ parameter can be sensitive to different brain/mental states, to brain/mind development, and to neural diversity, perhaps clarifying the boundaries between the normal and the ill mind.  Consistently, a key outcome of emergence of self-organized new states in complex systems is an adaptive behavior facing environmental constraints \cite{BassetGazzaniga2011}. Indeed, the concept of disease has also been related to reduced adaptive capabilities, and to the alteration of complexity \cite{Lauetal2022, Goldberger1997}. Along the lines of the seminal philosophical work of G. Canguilhem \cite{Canguilhem1991}, normality should be related to the ability to create new rules (i.e., adaptation) instead of living by the same old norms. We intend to further explore, in the future, the neural diversity through the most remarkable paradigm of complexity.

\section*{Acknowledgements}We acknowledge fruitful conversations with E.M.F. Curado, A.R. Plastino and R. Wedemann, as well as partial financial support by CNPq and Faperj.

\section*{Data Availability} 
The raw EEG wavesfrom all subjects are provided at data.mendeley.com under doi: 10.17632/dm3922zmpj.1

\section*{Additional Information} 
The authors declare no competing interests.

% \showacknow

\section*{Author contributions statement} 
D.M.A. design research, data analysis, , figure design, text writing and revision.
C.T. design research, mathematical formulation, text writing and revision.
H.S.L. mathematical formulation, data analysis, , figure design and text revision.


\begin{thebibliography}{99}
	
\bibitem{BassetGazzaniga2011} D.S. Bassett and M.S. Gazzaniga, {\it Understanding complexity in the human brain}, Trends Cogn. Sci. { \bf 15}(5), 200-9. (2011).
%~\url{ doi: 10.1016/j.tics.2011.03.006}.

\bibitem{Braslow2021}J.T. Braslow, J.S. Brekke and J. Levenson,  {\it Psychiatry's Myopia-Reclaiming the Social, Cultural, and Psychological in the Psychiatric Gaze.}, JAMA Psychiatry. {\bf 78} (4): 349-350 (2021).

\bibitem{Canguilhem1991} G. Canguilhem, {\it The Normal and the Pathological, trans. Carolyn R. Fawcett and Robert S. Cohen}, New York: Zone Books (1991).

\bibitem{Lauetal2022}Z.J. Lau, T. Pham, S.H.A. Chen and D. Makowski, { \it Brain entropy, fractal dimensions and predictability: A review of complexity measures for EEG in healthy and neuropsychiatric populations}, Eur J Neurosci, { \bf 56} (7):5047-5069 (2022). 
% doi: 10.1111/ejn.15800. 

\bibitem{Boltzmann1872}L. Boltzmann, {\it Weitere Studien \.uber das W\.armegleichgewicht unter Gas molek\.ulen} [{\it Further Studies on Thermal Equilibrium Between Gas Molecules}], Wien, Ber. {\bf 66}, 275 (1872).

\bibitem{Gibbs1901}J.W. Gibbs, {\it Elementary Principles in Statistical Mechanics -- Developed with Especial Reference to the Rational Foundation of Thermodynamics} (C. Scribner's Sons, New York, 1902; Yale University Press, New Haven (1948)); OX Bow Press, Woodbridge, Connecticut (1981)). 
%See also J.W. Gibbs, {\it The collected works},Vol.1, {\it Thermodynamics}, (Yale University Press, 1948).

\bibitem{Penrose1970}O. Penrose, {\it Foundations of Statistical Mechanics: A Deductive Treatment}, Pergamon, Oxford, page 167 (1970).

\bibitem{GellMannTsallis2004}M.  Gell-Mann and C. Tsallis, eds., {\it Nonextensive Entropy - Interdisciplinary Applications}, Oxford University Press, New York (2004).

\bibitem{Tsallis1988} C. Tsallis, {\it Possible generalization of Boltzmann-Gibbs statistics}, 
J. Stat. Phys. {\bf 52}, 479-487 (1988).

%\bibitem{TsallisMendesPlastino1998} C. Tsallis, R.S. Mendes and A.R. Plastino, {\it The role of constraints within generalized nonextensive statistics}, Physica A {\bf 261}, 534 (1998).

\bibitem{Tsallis2022} C. Tsallis, {\it Entropy}, Encyclopedia {\bf 2}, 264-300 (2022).

\bibitem{Tsallis2009}C. Tsallis, {\it Introduction to Nonextensive Statistical Mechanics--Approaching a Complex World}, Springer, New York (2009); Second Edition, Springer (2023). 

%\bibitem{TsallisCirto2013} C. Tsallis and L.J.L. Cirto, Eur. Phys. J. C 73 (2013) 2487.

\bibitem{YangZhen2021}H. Yang, J. Li and X. Zheng, { \it Different Influences of Negative and Neutral Emotional Interference on Working Memory in Trait Anxiety.} Front Psychol. {\bf 12}, 570552 (2021).
%~\url{ doi: 10.3389/fpsyg.2021.570552}.

\bibitem{Polich2006}J. Polich and J.R. Criado, { \it Neuropsychology and neuropharmacology of P3a and P3b}, Int J Psychophysiol. {\bf 60}(2), 172-85 (2006). 

\bibitem{Powell1970}M.J.D. Powell . \textit{A new algorithm for unconstrained optimization}, Nonlinear Programming. New York: Academic Press. pp. 31–66 (1970).

\bibitem{Coleman1996}T.F. Coleman and Y. Li,
\textit{An interior trust region approach for nonlinear minimization subject to bounds}, SIAM J. Optim. {\bf 6}, 418-445 (1996).

\bibitem{Askar2016}A. Askarzadeh,
\textit{A novel metaheuristic method for solving constrained engineering optimization problems: Crow search algorithm},
Computers $\&$ Structures {\bf 169}, 1-12 (2016).

\bibitem{Wildetal2023}R. Wild, M. Nötzold, M. Simpson,  T.D. Tran and R. Wester. { \it Tunnelling measured in a very slow ion–molecule reaction}. Nature (2023). https://doi.org/10.1038/s41586-023-05727-z

\bibitem{TsallisAnteneodoBorlandOsorio2003}C. Tsallis,  C. Anteneodo, L. Borland and R. Osorio, { \it Nonextensive statistical mechanics and economics}, Physica A: Statistical Mechanics and its Applications, {\bf 324}(1–2): 89-100 (2003). %~\url{https://doi.org/10.1016/S0378-4371(03)00042-6}.

\bibitem{TsallisUgur2020}C. Tsallis and U. Tirnakli, {\it Predicting COVID-19 Peaks Around the World}, Frontiers in Physics { \bf 8}, 2020.00217 (2020).
%~\url{ https://doi.org/10.3389/fphy.2020.00217}.  

\bibitem{Mitsokapasetal2021} E. Mitsokapas, B. Schäfer, R.J. Harris and C. Beck, { \it Statistical characterization of airplane delays}, Sci. Rep. { \bf 11},  7855 (2021). %~\url{https://doi.org/10.1038/s41598-021-87279-8}.

\bibitem{Dakhaleetal2023}B.J.  Dakhale, M. Sharma, M. Arif, K. Asthana, A.A. Bhurane, A.G. Kothari and U.R. Acharya,  {\it An automatic sleep-scoring system in elderly women with osteoporosis fractures using frequency localized finite orthogonal quadrature Fejer Korovkin Kernels},  Medical Engineering $\&$ Physics, { \bf 112}, 103956 (2023).
%~\url{https://dx.doi.org/10.1016/j.medengphy.2023.103956}. 

\bibitem{LudescherTsallisBunde2011}J. Ludescher, C. Tsallis and A. Bunde, {\it Universal behaviour of interoccurrence times between losses in financial markets: An analytical description}, Europhys. Lett. {\bf 95},  68002  (2011).

\bibitem{LutzRenzoni2013}E. Lutz and F. Renzoni, {\it Beyond Boltzmann-Gibbs statistical mechanics in optical lattices}, Nature Physics {\bf 9}, 615-619 (2013).

\bibitem{BogachevKayumovBunde2014}M.I. Bogachev, A.R. Kayumov and A. Bunde, {\it Universal internucleotide statistics in full genomes: A footprint of the DNA structure and packaging?}, PLoS ONE {\bf 9} (12), e112534 (2014).

\bibitem{CombeRichefeuStasiakAtman2015}G. Combe, V. Richefeu, M. Stasiak and A.P.F. Atman, {\it Experimental validation of nonextensive scaling law in confined granular media}, Phys. Rev. Lett. {\bf 115}, 238301 (2015).

\bibitem{GrecoTsallisRapisardaPluchinoFicheraContrafatto2020}A. Greco, C. Tsallis, A. Rapisarda, A. Pluchino, G. Fichera and L. Contrafatto, {\it  Acoustic emissions in compression of building materials: q-statistics enables the anticipation of the breakdown point}, European Physical Journal Special Topics {\bf 229} (5), 841-849 (2020).

\bibitem{Sauseng2010}P. Sauseng, B. Griesmayr, R. Freunberger and W.  Klimesch, { \it Control mechanisms in working memory: a possible function of EEG theta oscillations.}, Neurosci Biobehav Rev. { \bf 34}, 1015-1022. (2010).
%~\url{ doi: 10.1016/j.neubiorev.2009.12.006}.

\bibitem{Goldberger1997}A.L. Goldberger, { \it Fractal variability versus pathologic periodicity: complexity loss and stereotypy in disease}, Perspect. Biol.  Med. { \bf 40}(4), 543-61 (1997).
%~\url{ doi: 10.1353/pbm.1997.0063}.


\end{thebibliography}
\end{document}